# Pressure-induced competition between superconductivity and Kondo effect in CeFeAsO$_{1-x}$F$_x$ (x=0.16 and 0.3)

Liling Sun[1]*, Xi Dai[1], Chao Zhang[1], Wei Yi[1], Genfu Chen[1], Nanlin Wang[1], Lirong Zheng[2], Zheng Jiang[3], Xiangjun Wei[3], Yuying Huang[3], Jie Yang[1], Zhian Ren[1], Wei Lu[1], Xiaoli Dong[1], Guangcan Che[1], Qi Wu[1], Hong Ding[1], Jing Liu[2], Tiandou Hu[2] and Zhongxian Zhao[1]*

[1]Institute of Physics and Beijing National Laboratory for Condensed Matter Physics, Chinese Academy of Sciences, Beijing 100190, China;
[2]Beijing Synchrotron Radiation Facility，Institute of High Energy Physics, Chinese Academy of Sciences, Beijing 100039, China;
[3]Shanghai Synchrotron Radiation Facilities, Shanghai Institute of Applied Physics, Chinese Academy of Sciences，Shanghai 201204, China

We studied high-pressure behavior of CeFeAsO$_{1-x}$F$_x$ superconductors with x=0.16 and x=0.3 by *in-situ* measurements of electrical resistance, x-ray diffraction (XRD) and x-ray absorption spectroscopy (XAS) with diamond anvil cell (DAC). A pressure-induced quantum phase transition from the superconducting state to the non-superconducting Kondo screened phase associated with a volume collapse was discovered in the layered CeFeAsO$_{1-x}$F$_x$ compounds. The XAS data of Ce-L$_3$ in CeFeAsO$_{0.7}$F$_{0.3}$ clearly show a spectral weight transfer from the main line to the satellite line after the transition, demonstrating that Ce's valence changes under high pressure. Comprehensive experimental results and analysis in this paper provide some insight into the connection among superconductivity, valence change and structural phase transition, which reveals a picture of pressure-induced competition between Kondo singlet and BCS singlet in the Ce-pnictide superconductors.

Corresponding authors:
llsun@aphy.iphy.ac.cn;
zhxzhao@aphy.iphy.ac.cn.

Understanding the interplay between the RE 4f-electrons (RE is rare earth element) and itinerant electrons in RE-containing superconductors is one of important issues of current research in solid state physics. Generally, physical pressure, chemical substitution or doping influence the interaction between the 4f-electrons and itinerant electrons. The transition from an itinerant to a localized state may lead to the formation of different competing phases. The system of $CeFeAsO_{1-x}F_x$ is one of pnictide superconductors [1-2] which have a common layered crystal structure, belonging to the tetragonal *P4/mmm* space group at room temperature [2]. The itinerant electrons are confined to the FeAs structural units as conduction layers [3-4]. For the parent compound CeFeAsO, it is not superconducting but show an anomaly around 150 K in resistivity and magnetic susceptibility respectively [2]. Recent studies [5-6] show that this anomaly is caused by the SDW instability. Superconductivity emerges after suppression of the SDW state by fluorine doping. It is surprisingly unusual that the superconducting transition temperature (Tc= 41- 46 K) of $CeFeAsO_{1-x}F_x$ is much higher than that of conventional Ce-containing superconductors [7], which brings a great interest in understanding the physics in this new kind of superconductors.

Pressure can play an important role in the search for underlying physical mechanism of the competition among different quantum phases, because it can reduce interatomic distance and thus change the state of electrons [8-9]. Recent x-ray absorption and photoemission spectroscopy studies on $CeFeAsO_{1-x}F_x$ at ambient pressure show that Ce-4f electrons are localized at ~1.7 eV below the Fermi level [10]. This result implies that pressure may induce many novel phenomena such as valence transition and Kondo effect in these compounds.

Recently theoretical calculations on Ce-containing pnictide superconductors indicated that a competition between Kondo screening and superconductivity may take place when pressure is applied, which has been evidenced by an exponential increment of the Kondo temperature ($T_K$), and pointed out that the rapid increase in $T_K$ would destroy superconductivity due to Kondo screening effect [11-14]. For this prediction, there has not been an experimental observation of the actual pressure-induced full suppression of superconductivity in the Ce-containing pnictides to date. Here, we report an experimental finding of competition between superconductivity and Kondo effect in CeFeAsO$_{1-x}$F$_x$ through *in-situ* measurements of high-pressure electrical resistance, x-ray diffraction (XRD) and x-ray absorption spectroscopy (XAS). Our results indicate that the superconductivity of CeFeAsO$_{1-x}$F$_x$ (x=0.16 and 0.3) is suppressed at pressure where an iso-structure phase transition occurs. By performing high-pressure XAS experiments, we found that Ce-L$_3$ absorption edge showed a clear satellite structure after the iso-structural phase transition. Comparing with the high-pressure data of cerium metal, we propose that the pressure-induced iso-structural phase transition in CeFeAsO$_{1-x}$F$_x$ is of electronic origin. Namely it is driven by the Kondo screening effect between the 4f electron of cerium and the conduction bands mainly from the 3d shell of iron.

CeFeAsO$_{1-x}$F$_x$ sample with x=0.16 was synthesized by the solid reaction method at ambient pressure [2]. To obtain more fluorine doping, the sample with x=0.3 was synthesized under high pressure and high temperature [15]. The resulting samples were characterized by powder XRD with Cu Kα radiation at room temperature. A nearly single phase with a common ZrCuSiAs-type structure was achieved for the both samples. The lattice parameters for the x=0.16 and x=0.3 samples are a=3.989 Å, c=8.631 Å and a=3.985 Å, c=8.595 Å respectively.

High pressures were generated using diamond anvils. The anvils with 300 μm flat were employed for all measurements in this study. In the resistance measurements, the standard four-probe technique was adopted, in which four 2-μm-thick platinum leads are insulated from the preindented rhenium gasket by a thin layer of the mixture of cubic boron nitride and epoxy. Powder sample taken from a synthesized pellet was re-pressed into a flake and then the flake was loaded into a diamond anvil cell made of Be-Cu alloy. The superconductivity transition of the sample at each loading point was measured using a closed cycle refrigerator. High-pressure angle-dispersive XRD and high-pressure XAS experiments were carried out at room temperature at Beijing Synchrotron Radiation Facility and Shanghai Synchrotron Radiation Facility. In order to minimize x-ray absorption by the diamonds, partial perforated diamond anvils were used for XAS measurements in the transmission mode. The total thickness of the partial perforated anvils was reduced from 4.6 mm to 1 mm. The samples were loaded with the silicone fluid, because silicone fluid with initial viscosity of 1 cst can maintain the sample in a hydrostatic pressure environment up to 30 GPa [16]. Pressure was determined by ruby fluorescence [17].

We start from the electrical resistance and magnetization measurements for the samples investigated at ambient pressure. The temperature dependence of resistance and magnetization of the x=0.3 sample is shown in Fig. 1(a) and 1(b). It is seen that the sample has a bulk superconducting nature and its onset $Tc$ is ~46 K. While the onset $Tc$ of the x=0.16 sample is ~41 K [2], indicating that the more fluorine doping into the compound favors $Tc$ enhancement.

In Fig.2 (a) and (c), we present temperature (T) dependence of electrical resistance (R) of $CeFeAsO_{1-x}F_x$ (x=0.3 and 0.6) samples. It is found that the R-T curves of the both samples become broader and shift towards lower temperature with increasing pressure. Upon further uploading, the superconductivity is suppressed dramatically at 8.6 GPa for the x=0.3 sample and 9.6 GPa for the x=0.16 sample, disappears at 10 GPa and 12.8 GPa respectively. When downloading from the highest pressure, we found that superconductivity can be recovered, as shown in Fig.2 (b) and (d). We noted that the resistance of the samples studied did not display zero value under high pressure. To investigate the origin of nonzero resistance of the compressed samples, the microstructure was imaged by using a scanning electron microscopy. Many micro-cracks were observed from the sample recovered from compression, as shown in inset of Fig.2 (b), suggesting that the nonzero resistance background at low temperature is likely caused by these micro-cracks.

To extract more information of pressure effect on $Tc$, we plotted the onset $Tc$ of the two samples versus pressure in Fig.3. Here we define the onset $Tc$ as the temperature where $dR/dT$ rises rapidly. It is clearly seen that $Tcs$ of the two samples decrease gradually when pressure increases from ambient to ~8 GPa, drop remarkably at 8.6 GPa (x=0.3 sample) and 9.6 GPa (x=0.16 sample), and disappear at 10 GPa (x=0.3 sample) and 12.8 GPa (x=0.16 sample). For comparison, the results of x=0.12 sample measured by Zocco et al [18] was plotted in Fig.3. The negative pressure effect on $Tc$ in the three samples is in good agreement below 8 GPa. The rate of decrease in $Tc$, however, varies for the sample with different fluorine doping level. It is likely that the $Tc$ was suppressed more drastically in the sample with more fluorine doping. The negative effect of pressure on $Tc$ in $CeFeAsO_{1-y}$ compound was also observed by Takeshita et al [19], indicating that pressure has an equivalent effect on $Tc$ for Ce-containing pnictides either with fluorine doping or oxygen vacancy. To get insight into the role of Ce element in CeFeAsOF compounds, we compared pressure dependence of $Tc$ with $SmFeAsO_{0.85}$, $LaFeAsO_{0.5}F_{0.5}$ and $LaFeAsO_{0.89}F_{0.11}$ in the same pressure level [20-22]. $Tcs$ of (Sm, La)-containing superconductors decrease with pressure, as shown in inset of Fig.3, but are not suppressed to zero even at pressure up to 20 GPa, in sharp contrast to the high-pressure behavior of $CeFeAsO_{1-x}F_x$ and $CeFeAsO_{1-y}$ [19]. This suggests that cerium plays a special role for the $Tc$ disappearance in $CeFeAsO_{1-x}F_x$.

We consider three possibilities which may tightly account for the observed phenomena. First, pressure-induced structural phase transition needs to be verified. Several investigations have been shown that the $Tc$ can be suppressed when the crystal structure changes from one to the other [23-24]. Second, pressure-induced valence transition from $Ce^{3+}$ to $Ce^{4+}$ is an important factor because the transition may also suppress the $Tc$ due to electron overdoping. Third, Kondo effect caused by hybridization between the localized 4f electrons and itinerant electrons in Ce-containing compounds is a possible issue to destroy superconductivity since high pressure can lead to a strong enhancement of the hybridization.

To investigate whether the disappearance of $Tc$ in $CeFeAsO_{1-x}F_x$ is related to structural change, we performed *in-situ* XRD measurements for the x=0.16 and x=0.3 samples in a diamond anvil cell at room temperature. No new feature was observed in the diffraction patterns under pressure up to 21 GPa for both samples, indicating that the crystal structure of the samples remains unchanged, in another word, still stays in a tetragonal form at pressure where the samples lost their superconductivity. The XRD results rule out the first possibility mentioned above. Then we estimated the pressure dependence of the volume change according to the lattice parameters, as shown in Fig.4. A clear discontinuity in the volume-pressure curves are found in the two samples, demonstrating that a first-order iso-structural phase transition occurred under high pressure. The volume collapses by 2.6(1) % at 8 GPa for the x=0.16 sample and 2.1(1)% at 8.3 GPa for the x=0.3 sample. The presence of the volume collapse at pressure where the superconductivity is suppressed reflects a direct connection between them.

Further information was obtained by studying high-pressure XAS at room temperature. We performed four separate XAS measurements on the Ce-$L_3$ absorption edge for the x=0.3 sample. The pressure dependence of the Ce-$L_3$ edge of the sample is displayed in Fig.5 (a). The position of the $L_3$-edge does not change with increasing pressure. However, the intensity of main peak at 5.730 keV associated with $4f^1$ configuration has been suppressed when pressure is applied, while the intensity of a small satellite peak at 5.741 keV which has been attributed to the presence of the $4f^0$ configuration in the initial state appears increased at the same time. When the pressure is released from the maximum value to 4.5 GPa, the intensity of the main peak increases again whereas the intensity of the satellite decreases back to the low-pressure state, as seen in Fig.5 (b), consistent with our resistance results in the similar pressure range where the superconductivity of the sample is recovered. Then we carefully investigated the pressure dependence of the mean valence ($v$) of Ce ions by using a widely used method $v=3+I_{satellite}/(I_{main}+I_{satellite})$ [25-26], where $I$ represents amplitudes of spectral peak, and found that $v$ has a small change upon increasing pressure, varying from 3.0 at 1 GPa to 3.1 at 11.3 GPa.

Next, we compared our high-pressure XAS results with that of cerium metal. Figure 5(c) shows $L_3$-XAS data of cerium metal together with its pressure dependence of relative volume change [25, 27]. Note that high-pressure behavior in CeFeAsO$_{1-x}$F$_x$ resembles the γ-α phase transition under high pressure in cerium metal, in latter case a similar satellite peak at the same relative position to the main peak emerges only in the α-phase, revealing that the spectra weight transfer is tightly associated with the γ-α phase transition. XRD measurements on cerium metal showed that the volume collapses about 15% after iso-structural γ-α phase transition [27]. The mechanism of the γ-to-α transition in cerium metal has been debated in the literature for a long time. Several scenarios have been proposed to explain the electronic origin of this transition, including valence change, Mott transition, and Kondo effect etc. Among them, the Kondo Volume Collapse (KVC) scenario received more attention, and there are more and more evidences support KVC from both numerical simulation and experiments [25, 28-30]. The similarity of the iso-structural transition and spectra weight transfer of cerium ions in $CeFeAsO_{1-x}F_x$ to cerium metal implies that the KVC scenario may be applied to $CeFeAsO_{1-x}F_x$ system. We thus propose that the full suppression of superconductivity under high pressure in $CeFeAsO_{1-x}F_x$ is possibly caused by Kondo effect.

In KVC, the important difference between α phase and γ phase in cerium metal is their Kondo temperatures. Numerical simulation and experiments [27, 30-32] indicated that the Kondo temperature of cerium metal is above 1000 K in α phase and negligible in γ phase. Though our high-pressure XRD and XAS measurements were carried out at room temperature, we still can observe the similar behavior in $CeFeAsO_{1-x}F_x$ to cerium metal, we thus anticipate that the Kondo temperature of the $CeFeAsO_{1-x}F_x$ should be higher than room temperature.

In summary, pressure-induced full suppression of superconductivity in $CeFeAsO_{1-x}F$ (x=0.16 and 0.3) is found by *in-situ* resistance measurements with a diamond anvil cell. Based on the analysis of high-pressure XRD and XAS experimental results, we propose that the disappearance of Tc in $CeFeAsO_{1-x}F$ is due to a competition between superconducting phase and Kondo screened phase under high pressure. A clearer physical picture can be described as: the formation of the Kondo singlet between Ce local moments and Fe 3d electrons would break Cooper pairs in the FeAs layers, and then kill the superconductivity. Furthermore, our study reveals that doping concentration of fluorine in $CeFeAsO_{1-x}F$ influences the critical pressure of *Tc* disappearance, i.e. more fluorine doping shifts the critical pressure to lower side.

**Acknowledgements**

We sincerely thank J. Y. Ma and S. Q. Gu of SSRF for XAS experiment help. Authors wish to thank the National Science Foundation of China for its support of this research through Grant No. 10874230, 10874211, 10804127 and 11074294. This work was also supported by 973 project (2010CB923000) and Chinese Academy of Sciences.

**Figure legends**

Figure 1: (a) Electrical resistance as a function of temperature of the x=0.3 sample. (b) Temperature dependence of dc magnetization of the x=0.3 sample measured under 1 Oe after zero field cooling and field cooling respectively.

Figure 2: Temperature dependence of normalized resistance $R/R_{80K}$ of CeFeAsO$_{1-x}$F$_x$ samples at different pressures, showing data upon uploading (a), (c) and downloading (b), (d). SEM image of compressed sample with x=0.3 is displayed in inset (b).

Figure 3: Pressure dependence of *Tcs* of CeFeAsO$_{1-x}$F$_x$ (x=0.16 and 0.3) samples. The data of the x=0.12 sample is from Ref.18. The inset shows *P(Tc)* in SmFeAsO$_{0.85}$, LaFeAsO$_{0.5}$F$_{0.5}$ and LaFeAsO$_{0.89}$F$_{0.11}$.

Figure 4: Pressure dependence of unit cell volume for the x=0.16 and x=0.3 samples.

Figure 5: Ce-L$_3$ x-ray absorption spectrum of the x=0.3 sample (a) upon uploading and (b) downloading, and (c) XAS data of the $\gamma$ and $\alpha$ phase in cerium metal together with pressure dependence of atomic volume change.

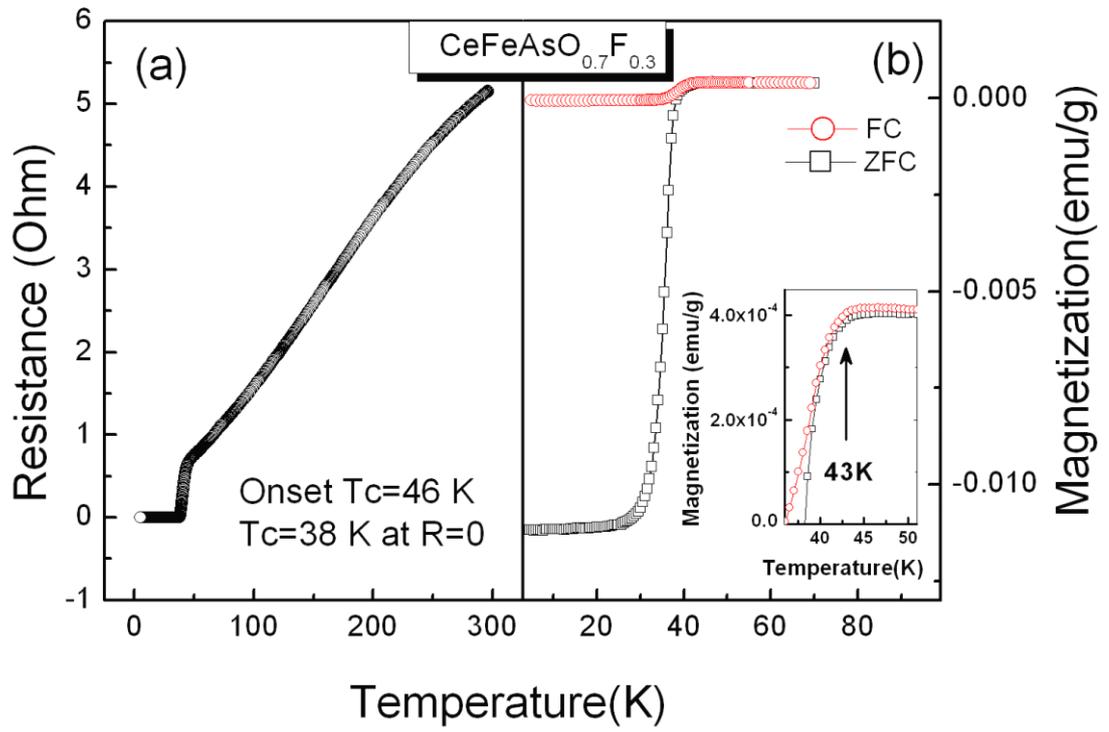

Fig.1　Sun et al

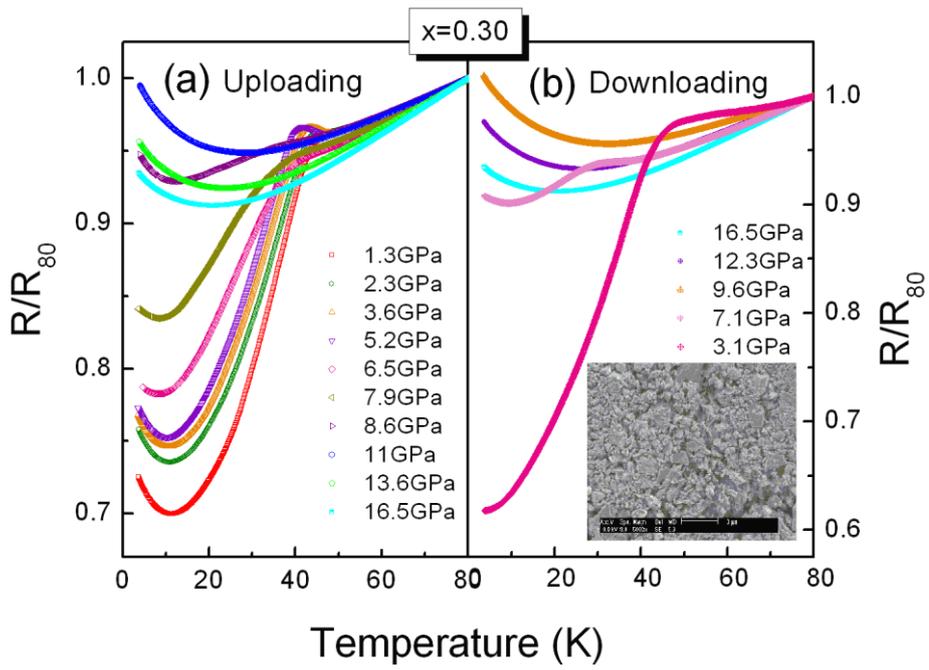

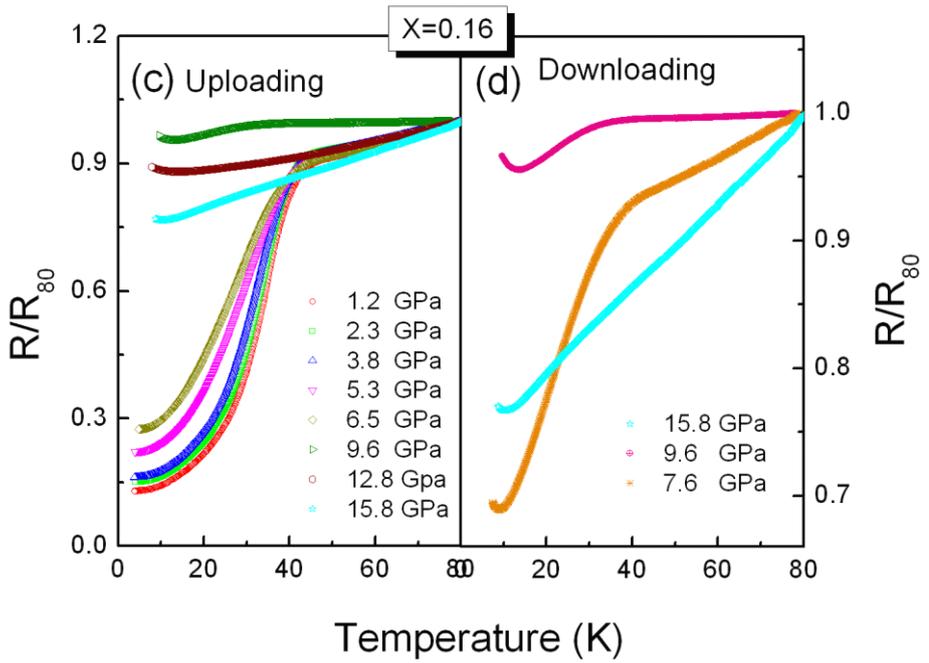

Fig.2  Sun et al

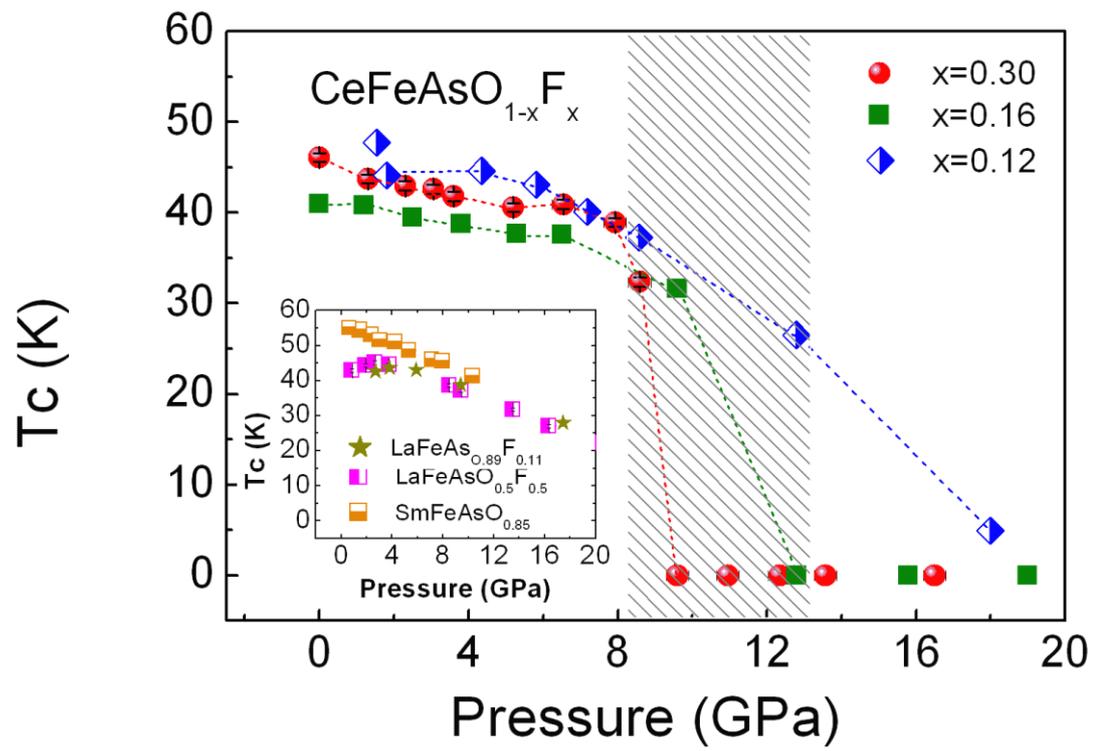

Fig.3    Sun et al

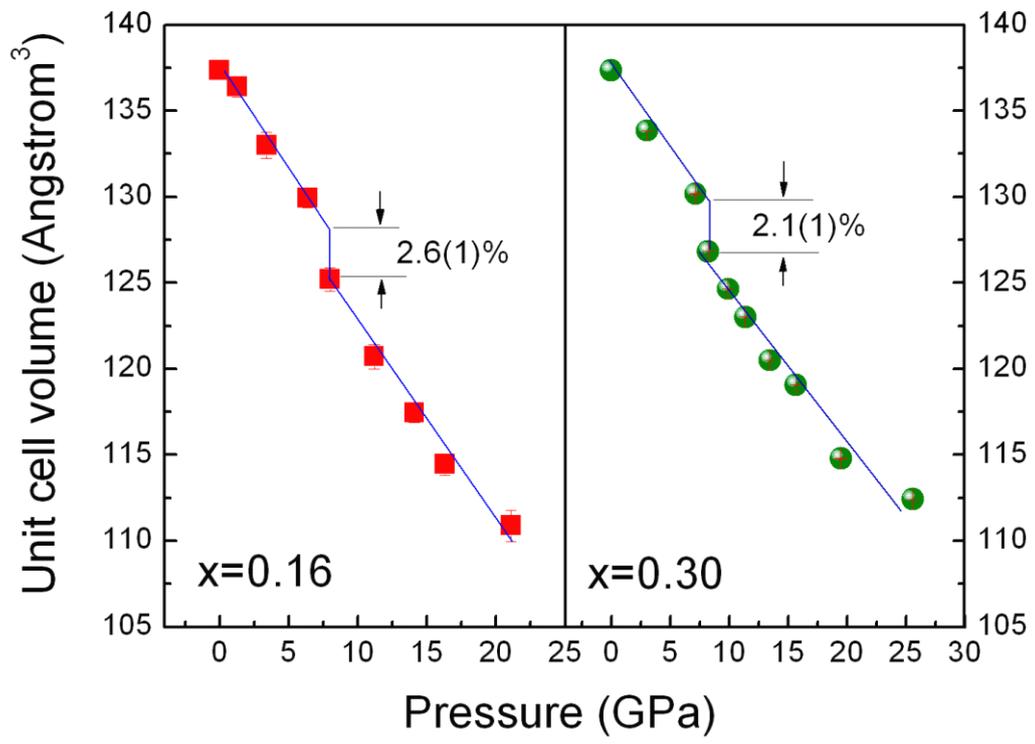

Fig.4    Sun et al

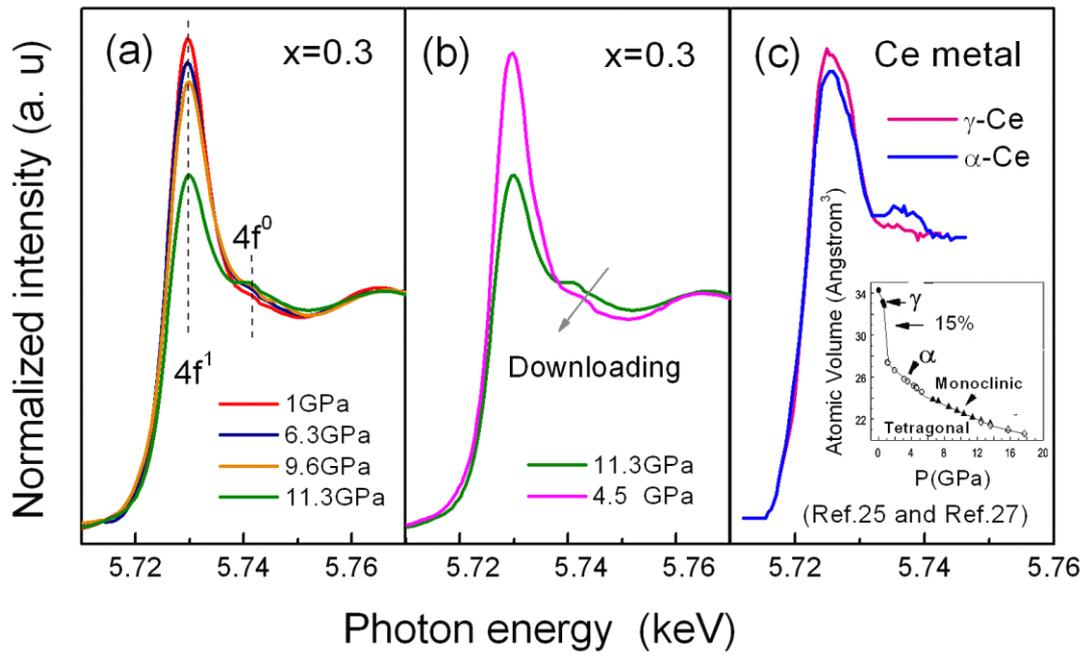

Fig.5    Sun et al